# Correctness Comparison of ChatGPT-4, Gemini, Claude-3, and Copilot for Spatial Tasks

Hartwig H. Hochmair[a]*, Levente Juhász[b], and Takoda Kemp[a]

[a]School of Forest, Fisheries, and Geomatics Sciences, Fort Lauderdale Research and Education Center, University of Florida, Davie, FL, U.S.A.; [b]GIS Center, Florida International University, Miami, FL, U.S.A.

**Abstract:** Generative AI including large language models (LLMs) has recently gained significant interest in the geo-science community through its versatile task-solving capabilities including programming, arithmetic reasoning, generation of sample data, time-series forecasting, toponym recognition, or image classification. Existing performance assessments of LLMs for spatial tasks have primarily focused on ChatGPT, whereas other chatbots received less attention. To narrow this research gap, this study conducts a zero-shot correctness evaluation for a set of 76 spatial tasks across seven task categories assigned to four prominent chatbots, i.e., ChatGPT-4, Gemini, Claude-3, and Copilot. The chatbots generally performed well on tasks related to spatial literacy, GIS theory, and interpretation of programming code and functions, but revealed weaknesses in mapping, code writing, and spatial reasoning. Furthermore, there was a significant difference in correctness of results between the four chatbots. Responses from repeated tasks assigned to each chatbot showed a high level of consistency in responses with matching rates of over 80% for most task categories in the four chatbots.

## Introduction

Large Language Models (LLMs), which simulate human-like conversations, have recently gained widespread popularity due to their versatility and use in conversational agents, text summarization, information retrieval, coding, computer science tasks, translation, reasoning, or solving arithmetic problems, among others. LLMs, such as OpenAI's ChatGPT, have performed well on academic and professional exams across various disciplines, such as mathematics, science, medicine, law (Kung et al., 2023; Ray, 2023; Rudolph, Tan, & Tan, 2023) and geographic information systems (Mooney, Cui, Guan, & Juhász, 2023). Despite these significant advancements, LLMs come with limitations and persistent challenges, including generating factually incorrect information due to hallucination or imperfect mathematical abilities including difficulties with unit conversions, handling numbers in scientific format, calculating descriptive statistics, or misplacing decimal points in arithmetic operations (Tyson, 2023). Their reliance on statistical patterns from training data also means that they lack human-like reasoning and understanding of context (Hadi et al., 2023). Other issues include limited generalizability, common sense, and domain-specific knowledge (Gu et al., 2021).

*Question benchmarks*

Recent initiatives compiled open datasets of spatial questions that can be used as benchmarks or gold standard for geospatial question-answering (GeoQA) machines. GeoQA machines, such as GeoQA2, take as input a question in natural language, work over geospatial knowledge graphs or geospatial datasets, such as DBpedia or OpenStreetMap, and retrieve knowledge using query languages, such as SPARQL and its geospatial extensions GeoSPARQL and stSPARQL (Punjani et al., 2023). A recently introduced GeoQA system called GeoQAMap first translates natural language questions into GeoSPARQL queries leveraging ChatGPT's natural language processing capabilities, then retrieves geospatial information from the Wikidata endpoint in JSON, and finally generates interactive maps using Python libraries (Feng, Ding, & Xiao, 2023).
  Examples of datasets for benchmarking GeoQA machines include GeoQuestions201 with 201 natural geospatial language questions (Punjani et al., 2018), GeoAnQu with 429 geo-analytic questions (H. Xu et al., 2020), and GeoQuestions1089 with 1089 questions and their answers over the geospatial knowledge graph YAGO2geo (Kefalidis et al., 2023). The questions in these datasets can be categorized (Punjani et al., 2018), such as those asking for the location of a feature or those determining the geospatial relation with another feature (e.g. east of). Whereas some questions can be useful for evaluating the response accuracy of LLMs, most of them are not suitable for this purpose as their answers depend on the dataset at hand and the analysis method applied, such as for "How many buildings are affected by a hurricane in Oleander?" Questions tend also to be subject to interpretation and the applied level of spatial and temporal granularity of analysis. Examples include "What is the travel behavior of individuals in the Tampa Bay region?" or "Which parks are near Trafalgar Square?". Travel behavior could be described for different time periods (leading to different answers), and what is meant by "near" is not generally agreed upon. This problem was previously described as *indirect QA* (Scheider, Nyamsuren, Kruiger, & Xu, 2021). Due to these problems, re-use of spatial questions and tasks from earlier chatbot assessments (Borji & Mohammadian, 2023; Cohn, 2023; Mooney et al., 2023) for a chatbot comparison study is a more promising and objective option.
  Recent work also examined the problem-solving capabilities of large multimodal models (LMMs) and introduced multimodal benchmark questions which cover various application disciplines, such as art and design, technology, engineering, or mathematics and include a wide range of image types, such as charts, diagrams, maps and music sheets (J. Li, Li, Savarese, & Hoi, 2023; Lu et al., 2024; Yue et al., 2023).

*LLM comparison*

ChatGPT-3.5 was the first major LLM-based chatbot launched on 30 November 2022, closely followed by Google's Bard (now Gemini), Bing Chat (now Copilot), GPT-4, and Anthropics' Claude-2 (Rudolph et al., 2023). ChatGPT is the most widely used chatbot in the world with over 100 million weekly active users and 1 billion visits every month, followed by Gemini with an average of around 331 million monthly visits (Shewale, 2023).
  The emergence of LLM-based chatbot alternatives to ChatGPT resulted in chatbot performance comparison studies. For example, the evaluation of 15 questions from different disciplines including sociology, business, mathematics, history, astronomy, and art history relevant to higher education revealed that GPT-4 performed the best, followed by GPT-3.5, Bing Chat, and Bard (Rudolph et al., 2023). The same ranking of chatbot performance was found when evaluating the quality of responses to myopia (near-sightedness) related queries (Lim et al., 2023). Another comparison study, which also included Claude, evaluated LLMs in 1002 questions encompassing 27 categories (Borji & Mohammadian, 2023). It demonstrated a success rate of 84.1% for GPT-4, 78.3% for GPT-3.5, 64.5 % for Claude, and 62.4% for Bard. Chatbots demonstrated proficiency in language understanding (spelling, grammar, translation, vocabulary, etc.) and facts, but encountered difficulties in mathematics, coding, and reasoning. Comparison of the performance of three LLMs on a benchmark set of 147 undergraduate-level control problems, which combine mathematical theory and engineering, revealed that Claude 3 Opus (58.5% baseline accuracy) outperformed GPT-4 (45.6%) and Gemini 1.0 Ultra: (34.0%) (Kevian et al., 2024). All three LLMs faced difficulties in handling problems involving visual elements such as Bode plots and Nyquist plots.

  A research gap identified in reviewing previous work is that the evaluation of LLMs on spatial tasks (e.g. spatial reasoning, spatial literacy, GIS operations, spatial data acquisition, toponym recognition, mapping, urban geography, time series forecasting) was primarily focused on OpenAI's chatbots, i.e., GPT-3.5 or GPT-4 (Cohn, 2023; Juhász, Mooney, Hochmair, & Guan, 2023; F. Li, Hogg, & Cohn, 2024; Mai et al., acc.; Mooney et al., 2023; Tao & Xu, 2023) with only few exceptions (Manvi et al., 2023; Yin, Li, & Goldberg, 2023). One study, for example, compared the spatial reasoning activities of GPT-3.5, GPT-4 and Claude-2 in the context of the tic-tac-toe game on 3x3 and 5x5 grids, demonstrating a relative superiority of Claude-2 over GPT-4 in the 5x5 scenario but a slightly higher performance of GPT-4 in the 3x3 grid scenario (Liga & Pasetto, 2023). A new benchmark for path planning tasks was used in (Aghzal, Plaku, & Yao, 2024) to evaluate LLMs including GPT-4 via different prompting methodologies as well as BART and T5 of various sizes via fine-tuning, showing promise of few-shot GPT-4 in spatial reasoning.

  Evaluation of 14 LMMs, including GPT-4V(ision) and Gemini, identified significant challenges in disciplines which present more complex visual data and require intricate reasoning, such as business and science (Yue et al., 2023). The proprietary GPT-4V achieved an accuracy of 55.7%, whereas open-source models, such as BLIP2-FLAN-T5-XXL achieved lower accuracies of approximately 34%. GPT-4V was also found to perform best in an evaluation of 12 prominent foundation models with MathVista, a benchmark designed to combine challenges from diverse mathematical and visual tasks, with an overall accuracy of 49.9%, outperforming Multimodal Bard by 15.1% (Lu et al., 2024). Solving geospatial tasks often requires multi-modal approaches (Mooney et al., 2023). Foundation models for GeoAI should therefore address these challenges by integrating various data types such as text, images, and spatial vectors to enhance performance and applicability across different geospatial domains (Mai et al., acc.). In fact, the new version of GPT-4V already demonstrated proficiency in executing fundamental map reading and analysis tasks (J. Xu & Tao, 2024).

*Research objectives*

This research aims to provide a closer insight into the correctness of responses from four chatbots for 76 tasks across seven spatial and GIS related categories, using quantitative and qualitative analyses. This will enhance findings from previous work which focused on individual LLMs in evaluating their spatial capabilities. Task definitions in our experiments use a zero-shot prompting approach and thus rely on the model knowledge gained from training data without providing additional examples or demonstrations as part of the prompt.

Each task is presented twice which allows subsequent quantification of consistency in the results. For two task categories verbosity of responses will be examined which allows to better understand which chatbot to consult if more elaborate responses with additional background information are preferred.

**Materials and Methods**

The study analyses responses to tasks and questions provided to the four analyzed chatbots in their chat interface. For each task a new chat was initiated to prevent use of context for solving subsequent tasks. Each task was given twice so that consistency between responses could be assessed. The experiment was conducted at the beginning of May 2024 using chatbots based on LLM versions that were available at that time.

*Task categories*

Tasks were divided into seven categories with a spatial context and chosen to evaluate factual knowledge (i.e. spatial literacy, GIS theory), visualization (mapping) and code writing capabilities, and interpretation (of functions and code examples) and spatial reasoning skills. The categories are drawn from earlier LLM evaluation studies (Borji & Mohammadian, 2023; F. Li et al., 2024; Tao & Xu, 2023) using a subset of categories that are suitable for spatial tasks and questions. Questions for some categories (e.g. GIS concepts, spatial reasoning) were adopted from other studies (Cohn, 2023; Mooney et al., 2023). For the spatial literacy category, we aimed to replicate the structure of some questions used in GeoQA systems (Punjani et al., 2018). For the other categories (e.g. coding, function interpretation), tasks were created anew or adopted and modified from other sources.

Each task category contains 10 given tasks except for the spatial literacy category with 16 tasks which cover a wide range of question for different geographic areas. Tasks were provided as text prompt so that LLM analysis capabilities could be assessed using a single input mode without the intermediate step of image analysis which is prone to introducing perceptual errors (Yue et al., 2023). The URL to download the complete set of 76 tasks with their responses for each chatbot is provided in the data availability statement.

*Spatial literacy*

This category contained 13 questions and 3 tasks on geographic knowledge related to points of interest, countries, highways, rivers, elevations, distances, temperatures, and population. Some questions and tasks involved the relationship between geographic features, such as topological (e.g., bordering countries, crossing of rivers and roads, islands), distance (e.g., closest cities), directional (e.g., cardinal directions between states) or order (e.g., sort cities by elevation or temperature). No instructions about the desired response length were given. Sample questions and tasks include:

- In which German city crosses A60 the Rhine river?
- Are more than 5 countries bordering the Baltic Sea?
- What is the single most important geographic commonality between Mauritius, Sardegna, Guernsey and Bali?
- Which two cities among Dallas, Atlanta, Memphis, and Oklahoma City are closest to each other considering great circle distances?

- In which order does the Danube flow through Bratislava, Budapest, Linz and Vienna?
- List the UNESCO world heritage sites in Oman.

*GIS concepts*

This category contains a selection of six true/false and four multiple choice questions. These questions were adopted from an earlier study which examined the knowledge of GIS theory of ChatGPT-3.5 and ChatGPT-4 (Mooney et al., 2023) based on GIS exam questions in a popular textbook for introductory GIS courses (Bolstad & Manson, 2022). The topics in this task category comprise fundamental concepts of mapping, spatial statistics, spatial interpolation, and coordinate transformations and computations. Examples include:

- A degree of longitude spans approximately 110574 meters at the Equator. How many meters are spanned by a second of longitude at the Equator? Choose between following answers: (a) 30.7, (b) 22.2, (c) 123
- Is the following statement true or false? A larger scale map covers less ground than a smaller scale map of the same physical size.
- What type of attribute is human population (the number of people) in a U.S. county data layer? Choose between following answers: (a) interval/ratio, (b) Nominal, (c) Ordinal

*Mapping*

LLMs can leverage existing programming language libraries for map design and create executable programming code. They can also generate a URL that points towards a third-party online mapping service. In both cases it is necessary for an LLM to provide correct parameter information and follow syntax rules. In this study, chatbot capabilities to generate maps leveraging Mapbox resources (four tasks), Python libraries (three tasks) and R packages (three tasks) were examined. The Mapbox related set of tasks involved the Mapbox Static Images API and the Mapbox GL JS client-side JavaScript library. The following list provides one example from each task group:

- Create a Mapbox map link of Vienna with a marker at (16.3692, 48.2034)
- Generate code based on the matplotlib library which creates a map that shows the population and location of the 5 largest cities in the U.S.
- Generate code that uses the R tidyverse package to plot a world map

*Function interpretation*

This task category explores chatbot capabilities to explain the purpose of functions commonly used in the spatial sciences. Named labels, such as "distance" or "NDVI", were replaced with more generic labels (such as "a", "b"). This forced the LLM to solve a given task through analyzing the function structure rather than names. The ten functions given evolve around coordinate transformation, distance computation, spherical trigonometry, multispectral image analysis, and spatial regression. Two tasks (identify Epanechnikov kernel and spatial lag model) are as follows:

- Provide a term commonly used for this equation: $K(x) = 3/4 (1-x^2)$ for $|x| \leq 1$

- Explain in one sentence what this equation is called: $y=\rho Wy+X\beta+\varepsilon$

*Code explanation*

This task category is similar to the previous one except that the input is a programming algorithm expressed as Python code (five tasks) and R code (fives tasks). Some provided code snippets use programming libraries and packages, such as ArcPy, math, pathlib (Python) or sf (R). As before, function and variables names were replaced with generic labels. The provided Python code snippets include among others algorithms which create a binary slope map from an elevation raster, compute the shortest path using Dijkstra's algorithm, and find the point of intersection of two lines. The provided R code snippets compute among others the centroid of a polygon, produce a raster grid that overlaps with a given polygon provided as a shapefile, or find the X, Y tile coordinates in the web map tile system based on a given point location. The task associated with the latter algorithm is provided as an example below:

> Explain in one sentence the purpose of this function:
> ```
> my_calc <- function(lat, lon, z){
>   l <- lat * pi /180
>   n <- 2.0 ^ z
>   x <- floor((lon + 180.0) / 360.0 * n)
>   y = floor((1.0 - log(tan(l) + (1 / cos(l))) / pi) /
>       2.0 * n)
>   return(c(x, y))
> }
>
> my_calc(20, -80, 15)
> ```

*Coding*

This task category explores chatbot capabilities to write, modify, and complete programming code (four tasks in Python and two tasks in R), and to translate programming code from Python to R (two tasks) and from R to Python (two tasks). Some tasks involve handling spatial programming libraries and packages, such as ArcPy, shapely (Python), spatstat or sp (R). The six code writing/modification/completion tasks include the following:

- Generate Python code which: (1) sorts a set of points with given geographic coordinates from East to West; (2) uses the ArcPy library to generate a line geometry from a set of points with given geographic coordinates and saves the result as a shapefile
- Enhance provided Python code with the following functionality: (3) print the id (memory address) of local and global variables; (4) update values for two additional fields of the specified record in a given point feature class
- Complete the provided R code to: (5) plot three polygons based on given x/y point coordinates; (6) create and plot a raster map using indirect distance weighting with a power of 0.5

As an example, the input for task (5) is as follows:
> Complete this code to plot three polygons.
> ```
> install.packages("sp")
> ```

```
library(sp)
cs1 <- rbind(c(7, 5), c(10, 5), c(10, 0), c(5, 0))
cs2 <- rbind(c(5, 5), c(10, 5), c(9, 8))
cs3 <- rbind(c(7, 5), c(3, 5), c(5, 0))
```

Code translation tasks involve various challenges, such as handling spatial reference information and recursive functions. Following code algorithms needed to be translated:

- From R to Python: (1) generate polygons from points with x/y coordinates and plot them; (2) compute the Haversine distance between two points with given geographic coordinates
- From Python to R: (3) sort given point coordinates by longitude and save them as points in a shapefile; (4) sort numbers using a Quicksort algorithm

*Spatial reasoning*

Tasks in this category were primarily drawn from previous studies on spatial reasoning (Borji & Mohammadian, 2023; Cohn, 2023). Upon initial examination of explanations provided in chatbot responses we noticed that the reasoning behind some tasks was incorrect but still resulted in the correct answer due to the lack of specificity of the question. To avoid these situations, some tasks were either re-written to require provision of a solution (as opposed to only asking a question about the solution) or formulated as a multiple-choice question with a smaller chance of randomly picking the correct answer. For example, the question "Can you place six X's on a Tic Tac Toe board without making three-in-a-row in any direction?" was re-written as "Place six X's on a Tic Tac Toe board without making three-in-a-row in any direction".
    Tasks in this category required the identification of spatial relationships between objects (e.g. boxes or persons) based on a set of relative position descriptions, point connection and arrangement tasks on grid-like "worlds" (see Tic Tac Toe example above), and qualitative spatial reasoning tasks on the region connection calculus 8 (RCC-8). For the latter, three tasks were chosen from (Cohn, 2023), two of which were previously correctly answered in ChatGPT-4 in that study. One example of the qualitative spatial reasoning tasks on the RCC-8 reads as follows (some parts removed for brevity):

- Consider the following set of eight pairwise disjoint and mutually exhaustive binary spatial relations. These relations form part of the well-known RCC-8 qualitative spatial reasoning calculus. DC(x,y) means that x and y are disconnected and share no spatial parts. EC(x,y) means that x and y touch at a boundary but do not share any interior parts […]. TPP(x,y) means that x is part of y and touches y's boundary. TPPi(x,y) is the same as TPP(y,x). […]. If DC(x,y) and TPPi(y,z) then what are the possible relationships between x and z?

*Analysis methods*

The correctness of chatbot responses was analyzed both statistically and qualitatively. A task was counted as completed correctly if the responses in both attempts were correct. Open-ended tasks, such as code and function interpretation were considered correct if the relevant technical terms and essential behavior were present in the answer. URLs and programming code generated in response to mapping and coding tasks were tested in a Web browser (URL, HTML) or an integrated development environment, i.e., PyCharm (Python) and RStudio (R), respectively.

Besides a summary table which reports the percentage of correctly completed tasks in the seven task categories for the four analyzed chatbots, statistical significance of differences in the proportion of correct results between chatbots and between task categories were analyzed using a two-tailed Pearson's chi-square test. This was followed by post-hoc tests which adjust the significance level ($\alpha$) for multiple testing using the Bonferroni correction method to determine statistically significant differences in percent correct between individual pairs of chatbots and task categories.

Difference in the length of responses between spatial literacy tasks and GIS concepts tasks were analyzed using a Wilcoxon rank sum test. Differences in response length between the four chatbots for each of these two task categories were analyzed using a Kruskal-Wallis test, followed by a Dunn post-hoc test which compared response lengths between individual pairs of chatbots. Correctness outcomes of repeated tasks were used to compute the consistency (matching) of responses. Matching rates were summarized for chatbots and task categories.

Qualitative analysis of results discussed observed difficulties and challenges in the completion of tasks together with illustrations and examples.

**Results**

*Statistical analysis*

Table 1 summarizes the number of correctly completed tasks for each task category (rows) and chatbot (columns), with percentage correct shown in parentheses. Numbers in boldface highlight for each task category the chatbots with the highest percentage of correct responses, which are GPT-4 for five out of seven categories (some ties with Claude-3 and Copilot), Copilot for coding tasks, and both Gemini and Copilot for spatial reasoning. Column totals reveal that GPT-4 completes most tasks correctly (76.3%), whereas Gemini completes fewest tasks (55.3%) correctly. This pattern is also reflected in Figure 1a.

Percent correct varies between task categories (see Category total column in Table 1 and Figure 1b). It is highest for tasks related to GIS concepts (95.0%) and code explanation (95.0%), and lowest for mapping tasks (25.0%), followed by spatial reasoning tasks (32.5%).

[insert Table 1 about here]

Figure 1. Percentage of correct task completion grouped by chatbot (a) and task category (b). Pairs of tasks with significant differences in percent correct are connected through horizontal bars in (b).

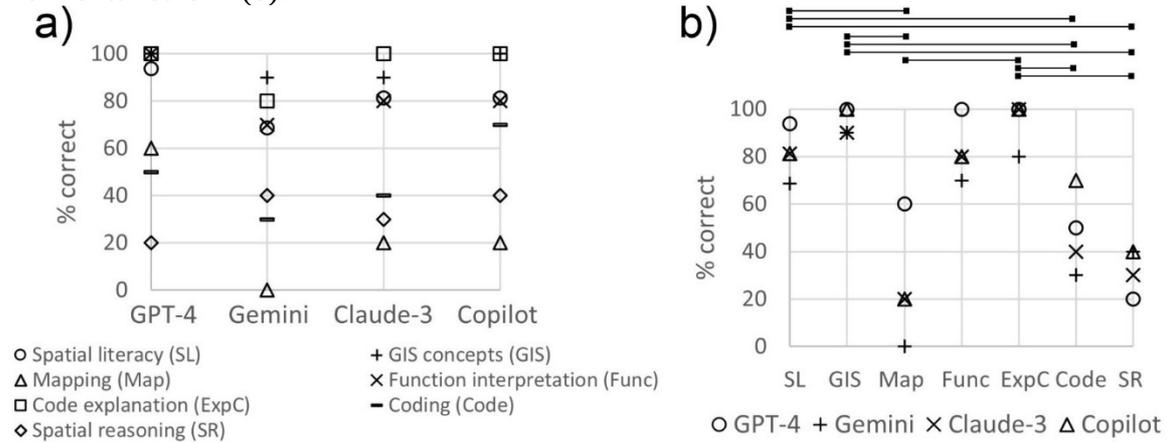

A chi-square test of independence was performed to examine the relation between chatbot and response correctness. Results show that there is a significant relationship between the two variables, $X^2$ (3, N = 304) = 8.47, p = 0.037. Post hoc comparisons were conducted using pairwise chi-square tests of independence between chatbots and response correctness applying a Bonferroni correction for multiple testing. Assuming a p-value of less than 0.05 to be considered statistically significant, an adjusted alpha level of .008 (0.05/6) was used for this purpose. Using the adjusted alpha level threshold of 0.008, no difference between any pair of chatbots was statistically significant, although the p value for the difference between GPT-4 and Gemini was close to that threshold, $X^2$ (1, N = 152) = 6.577, p = 0.010.

A chi-square test of independence showed that there was a significant association between task category and response correctness, $X^2$ (6, N =304) = 98.6, p < 0.001. Post hoc comparisons, which applied a Bonferroni correction for multiple testing with an adjusted alpha level of .0024 (0.05/21) for individual chi-square tests showed that differences were significant for 12 task pairs which reveal a distinct pattern (Table 2). That is, tasks related to spatial literacy, GIS concepts, function interpretation and code explanation resulted significantly more often in correct responses than mapping, coding, and spatial reasoning tasks. The task category pairs with significant correctness differences are visualized as horizontal bars in Figure 1b.

*[insert Table 2 about here]*

Figure 2a and b plot the percentage of matching correctness results in response to repeated tasks aggregated by task category for each chatbot. Consistency was 80% or higher in 25 out of 28 task categories evaluated across the four chatbots. Only mapping tasks for Claude-3 (60%) and Copilot (60%) and spatial reasoning tasks for GPT-4 (70%) had lower consistency rates, meaning that 4 out of 10 repeated mapping tasks for Claude-3 and Copilot and 3 out of 10 repeated spatial reasoning tasks for GPT-4 yielded different outcomes.

Figure 2. Consistency of responses for task categories, grouped by chatbot (a) and task category (b).

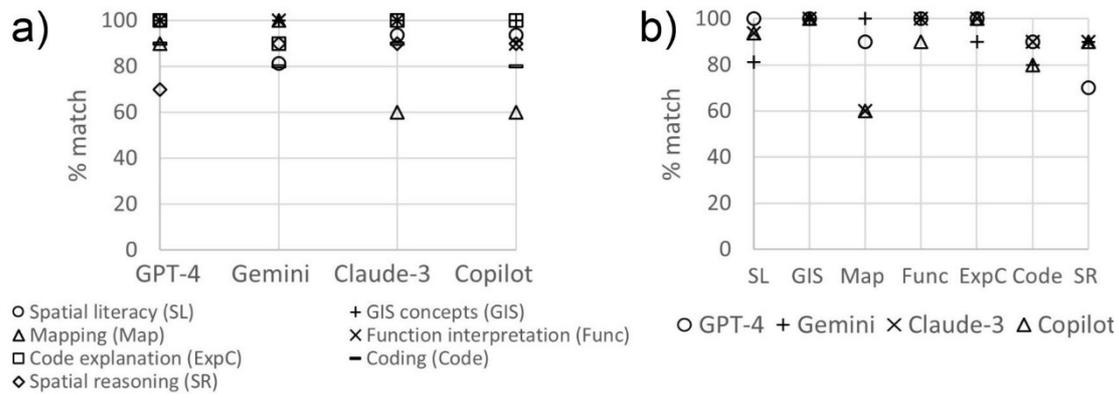

Table 3 lists the mean and median number of words in responses to the spatial literacy tasks (top half) and GIS concept tasks (bottom half) for the four compared chatbots, revealing that GPT-4 and Gemini provide most concise and Claude-3 provides most verbose answers, with Copilot in-between. For Copilot, which offers three conversation styles, the "more precise" option, which tends to provide shorter answers, was chosen for this study. A Wilcoxon rank sum test with continuity correction indicated that output was significantly longer for GIS concept tasks (MD = 90.0) than for spatial literacy tasks (MD = 59.3), W = 760, $p < 0.001$, demonstrating that the question category affects response length.

[insert Table 3 about here]

A Kruskal-Wallis test showed that the difference in central tendency of word counts between the four compared chatbots was significant both for spatial literacy tasks, $H(3, n = 64) = 9.632$, $p = .022$, and GIS concepts tasks, $H(3, n = 40) = 12.666$, $p = .005$. Post hoc comparisons using Dunn's method with a Bonferroni correction for multiple tests with an adjusted alpha level of .008 (0.05/6) showed a significant difference
- between GPT-4 and Claude-3, $p = 0.007$, for spatial literacy questions, and
- between GPT-4 and Claude-3, $p = 0.002$, and Gemini and Claude-3, $p = 0.002$, for GIS concepts questions.

These three chatbot pairs with significant differences in response lengths are annotated with small superscript letters in Table 3.

*Qualitative analysis*

*Spatial literacy*

Most tasks in this category (12 out of 16) were completed correctly by at least three chatbots. The four remaining tasks which revealed some challenges were the following: (1) Identify the German city in which the A60 intersects the Rhine river; (2) List the order in which the Danube flows through Bratislava, Budapest, Linz and Vienna; (3) Determine the cities of 250,000 or more residents within a 100 mile drive from Lexington, KY; (4) List the UNESCO world heritage sites in Oman. For the latter task both GPT-4 and Claude-3 missed one UNESCO world heritage site in Oman, i.e., Ancient City of Qalhat, which was added in 2018, long before these LLMs were trained. The reason for this omission is therefore unclear. For task (1) the incorrect result is based on factual errors as chatbot explanations in Gemini and Copilot revealed by stating that the A60 highway in Germany does not cross the Rhine River. In addition, Copilot demonstrated problems in sorting cities by average daytime temperature in July although it reported the correct temperature for each city.

*GIS concepts*

The only question which was not correctly answered by all chatbots was if grid north is in the direction of the north pole. Both Gemini and Claude-3 provided an incorrect response in both attempts.

*Mapping*

The tasks in this category proved to be the most challenging ones. GPT-4 was the only chatbot which completed more than half of the tasks correctly (6/10) and which also successfully handled at least one Mapbox mapping task in repeated attempts. That successful Mapbox task was the most complex one ("Create a Mapbox map with a line from Vienna to Munich") since it required generating an HTML script that integrated the Mapbox GL JS client-side JavaScript library (Figure 3a). None of the chatbots successfully completed the task to generate code that uses the R tmap package to map worldwide cities. Other Python libraries (e.g. matplotlib, pandas) and R packages (e.g., tmap) were more successfully applied to generate maps and scatterplots of largest cities in the U.S. (Figure 3b and c) and world maps (Figure 3d-f). No specific instructions about the map design were provided in the prompt. Therefore, produced map layouts do not necessarily meet strict cartographic mapping guidelines. Gemini was the only chatbot which failed in completing all 10 tasks.

Figure 3. Selected results of mapping tasks. GPT-4: Mapbox map with a line from Vienna to Munich (a); GPT-4: map showing the locations of the 5 largest cities in the U.S. using matplotlib library (b); GPT-4: scatter plot showing the locations of the 5 largest cities in the U.S. using pandas library (c); Claude-3: World map with orthographic map projection using matplotlib library (d); Claude-3: World map using R tmap package (e); Copilot: World map using R tidyverse package (f).

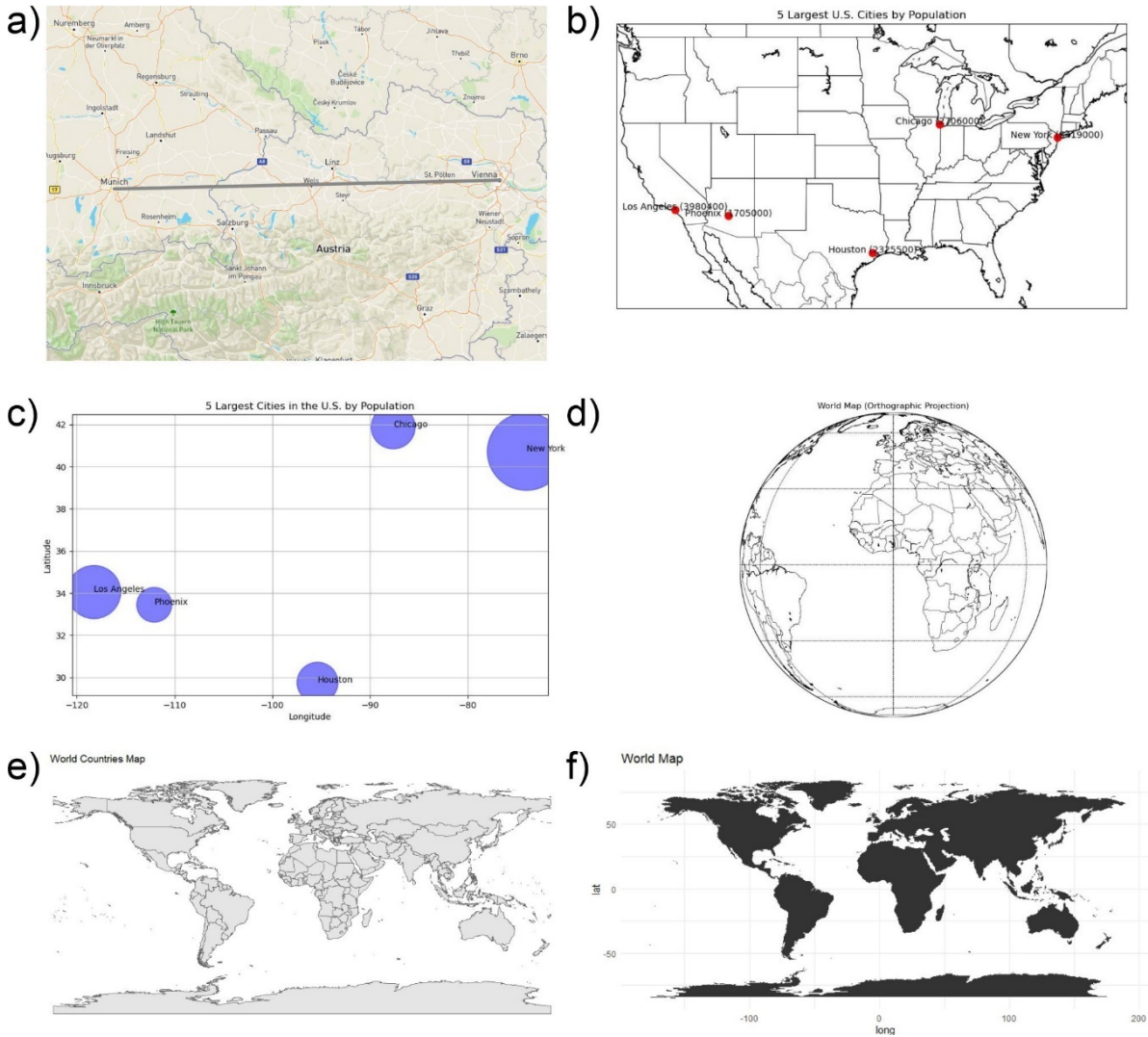

*Function interpretation*

GPT-4 identified and labelled all 10 provided functions correctly, with Claude-3 (8/10), Copilot (8/10), and Gemini (7/10) following closely. The only function which was not correctly recognized by two chatbots at the same time (Gemini and Copilot) was that for a semi-variogram (Equation 1):

$$g(h) = \frac{1}{2} E(Z(s) - Z(s+h))^2 \qquad (1)$$

Functions captioning 2-D Euclidean distance, NDVI as well as great circle distance and Law of Cosines in spherical trigonometry were correctly identified by all chatbots in both attempts. Equations for affine transformation, spatial lag model, Epanechnikov kernel, Mercator Projection and spherical excess were identified correctly by three chatbots.

*Code explanation*

Whereas GPT-4, Claude-3, and Copilot provided correct explanations for all 10 code snippets, Gemini failed to do so for two tasks (find point of intersection of two lines in

Python, find the centroid of a polygon in R). For example, for the line intersection task, Gemini describes the purpose of the code too generally as code which "solves a system of two linear equations using Cramer's rule".

*Coding*

The 10 tasks in this category proved to be challenging overall, with Copilot performing strongest (7/10) and Gemini performing poorest (3/10). The two code modification tasks in Python and the translation of the Haversine distance code snippet from R to Python are the only three coding tasks which were successfully completed in all four chatbots. As opposed to this, all four chatbots failed to translate a Quicksort algorithm implementation from Python to R.

Three chatbots (GPT-4, Claude-3, Copilot) managed the task to complete R code to plot three polygons. Results varied slightly between GPT-4 (Figure 4) and Claude-3/Copilot (Figure 5b).

Figure 6. Selected results of coding tasks: Complete R code to plot three polygons in GPT-4 (a) and Claude-3 and Copilot (b); Complete R code to plot a raster map using indirect distance weighting in Copilot (c).

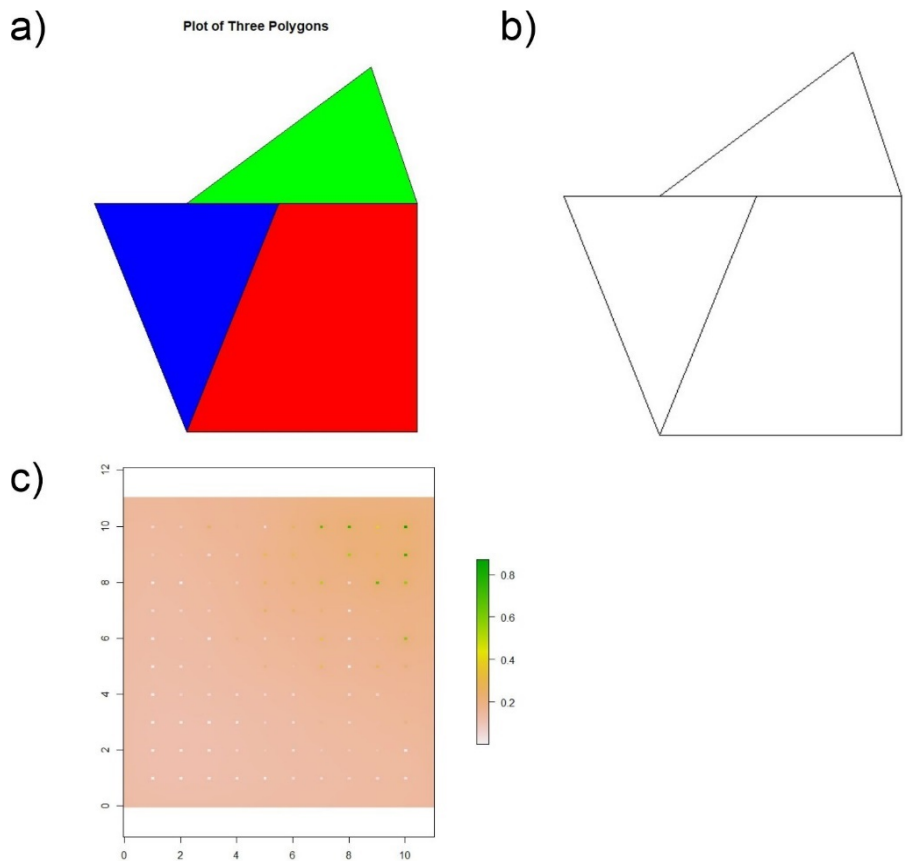

Copilot was the only chatbot to successfully complete the code snippet shown below to create and plot a raster map using indirect distance weighting with a power of 0.5 (Figure 7c). It was also the only chatbot generating Python code that correctly sorted three points with their given geographic coordinates from East to West. As opposed to this, the remaining three chatbots incorrectly sorted points from West to East.

```
library(spatstat)
```

```
        library(sp)

        xlist <- 1:10
        ylist <- 1:10
        comb <- expand.grid(xlist,ylist)
        labels <- runif(100)*(comb$Var1/10)*(comb$Var2/10)
        w = owin(c(0,11),c(0,11))
        ppp_rd<-ppp(comb$Var1,comb$Var2, marks=labels,window=w)
```

*Spatial reasoning*

Correctness was low for this task category with a range between 2/10 (GPT-4) and 4/10 (Gemini, Copilot). The only question correctly answered by all four chatbots was if two circles that both have a radius of 5 and whose distance between their centres is 9 intersect. All four chatbots failed in the three following tasks: (1) Provide the order of three cubic boxes (A, B, C) from top to bottom if C is immediately below A, and B is higher up than C; (2) Place six X's on a Tic Tac Toe board without making three-in-a-row in any direction; (3) RCC-8 qualitative spatial reasoning calculus: If NTPP(x,y) and TPP(y,z) then what are the possible relationships between x and z? GPT-4 was unable to solve any of the three RCC-8 tasks, whereas Gemini and Claude-3 successfully completed two of them and Copilot one, respectively.

**Discussion**

*Chatbot performance metrics*

Results presented in this study are a reflection of LLMs' current capabilities. Recent benchmark studies that compare GPT-3.5 and GPT-4 (Ali et al., 2023; Koubaa, 2023) indicate that the capabilities of LLMs improve rapidly. This was already shown in the case of an introductory GIS exam between these two variants of ChatGPT (Mooney et al., 2023). Thus, we expect that the performance related to spatial tasks will also improve over time. However, it must be noted that it is difficult to predict how LLMs will evolve, especially when it comes to their capabilities in spatial reasoning.

     Previous studies found GPT-4's performance to be superior compared to that of other chatbots in different disciplines such as business, mathematics, history, medicine, reasoning and language (Borji & Mohammadian, 2023; Lim et al., 2023; Rudolph et al., 2023). Our study adds a spatial focus to chatbot evaluations and reveals a more nuanced picture. That is, in line with previous findings, GPT-4 performed also best with spatial tasks and handled more tasks correctly (58/76) than Copilot (54/76), Claude-3 (49/76) and Gemini (42/76). Whereas the difference in percent correct between the chatbots was statistically significant (p = 0.037), comparison of percent correct between individual pairs of chatbots did not show any significant difference in post-hoc tests. Although GPT-4 was the single best performer in three task categories (spatial literacy, mapping, function interpretation) it performed poorest in spatial reasoning, and delivered also more erroneous results than Copilot in coding tasks.

     Analyzing results across all four chatbots revealed that tasks on spatial literacy, GIS concepts, function interpretation and code explanation were significantly more often correct than mapping, coding, and spatial reasoning tasks. Mapping tasks revealed that chatbots

faced difficulties in using the Mapbox APIs, with parameter values missing or incorrectly set. Similarly, coding tasks revealed problems in using functions from libraries or packages due to incorrect use of function parameters. Earlier studies showed that LLMs struggle with coding questions (Borji & Mohammadian, 2023), and that Bard (now Gemini) was the most erroneous performer in this category. This matches our findings. Despite these flaws and even occasional problems in running their own code, LLMs are commonly used as code-writing assistant (Stokel-Walker & Noorden, 2023) for tasks such as creating generic functions.

Spatial reasoning tasks were found to be challenging for LLMs in other experiments as well although they do better than chance (Cohn, 2023). Though LLMs can learn spatial concepts from text (Abdou et al., 2021) and possess a degree of abstract reasoning skills, they demonstrate weaknesses in spatial reasoning and planning. While current LLMs may possess some abstract task-solving skills, they often also rely on narrow, non-transferable procedures for task-solving (Wu et al., 2023). Similarly, investigation of the trustworthiness of ChatGPT and GPT-4 showed that these models frequently fall short of generating logically consistent predictions as measured by semantic, negation, symmetric, and transitive consistency, and that they also exhibit high levels of self-contradiction (Jang & Lukasiewicz, 2023).

The tasks in the spatial literacy and GIS concepts task categories did not stipulate a certain response length and thus allowed us to compare the verbosity of responses between chatbots and between task categories. Findings do partially support previous work which showed that GPT-4 provided shorter answers than Claude (Borji & Mohammadian, 2023). Hence, Claude-3 might be more suitable for chatbot users who are interested in obtaining more background information related to a task or question. While Bing Chat (now Copilot) was shown to provide brief answers compared to GPT-4 and Bard (Gemini) (Rudolph et al., 2023) this pattern was not observed in our experiments. That is, Copilot answers were longer than those of GPT-4 and Gemini, but this difference was not statistically significant. An earlier study (Scheider, Bartholomeus, & Verstegen, 2023) had 41 university teachers evaluate ChatGPT-3.5 responses to Geography and GIScience related questions along four quality scores. Results showed that 80% of answers had correctness scores, and 74% had completeness and clarity scores that would allow ChatGPT to pass the exam. As opposed to this, conciseness scores were much lower at 47% pointing towards overly long answers. While correctness, completeness, and conciseness were to some extent evaluated in our study setup, assessment of response clarity might be considered for chatbot evaluations in future work as well.

*Chatbot architecture and performance*

All chatbots included in this study are based on decoder-only transformers (Vaswani et al., 2017). While their general architecture is similar, each chatbot and their respective LLMs have distinct design philosophies, training datasets and optimization strategies, which contribute to their strengths and weaknesses in various task categories presented in this paper. Table 4 lists some details about these chatbots. ChatGPT-4 (OpenAI, 2023), Claude-3 (Anthropic, 2024) and Copilot are based on generative pre-trained transformer (GPT) models originally developed by OpenAI (Radford, Narasimhan, Salimans, & Sutskever, 2018).

[Insert Table 4 about here]

These models face challenges in spatial reasoning and advanced mapping tasks, which originates from their design and purpose. Due to their commercial nature, the exact architecture, training dataset and other technical details are not always available, which is for example the case with GPT-4 where the technical report published by its creators focuses on its capabilities (OpenAI, 2023). This closed structure makes it challenging to assess the effect of model architecture on performance. A detailed discussion of this association is therefore inherently speculative as it can only draw from industry news and user experience. For example, both ChatGPT and Copilot are based on the same model variant, however their performance across task categories is different. Two potential contributing factors are 1) the different setup of the chat interface that could use the models with different parameters (e.g. token length, temperature, etc.), and 2) Microsoft's access to a wider ecosystem, including emails, calendars, and web searches. Also, Microsoft acquired GitHub, a widely popular open-source code repository and version control system, which provides Copilot more training data for coding and programming related tasks. This might explain why Copilot performed best in coding related tasks (Table 1). To gain a deeper technical understanding of the relationship between each LLM and its spatial capabilities, more targeted approaches and surveys are needed, which are beyond the scope of this paper. However, our results demonstrate varied performance across different spatial task categories, highlighting the need to optimize LLM architectures for specific GeoAI applications. To improve the capabilities of future chatbots in managing complex spatial tasks we offer the following suggestions:

- Multimodal integration: Incorporate additional data types such as images, maps, and spatial vectors.
- Specialized training: Fine-tune models on domain-specific datasets that include geospatial information and spatial reasoning tasks.
- Advanced prompt engineering: Utilize techniques such as chain-of-thought and self-checking prompts to improve reasoning and decision-making capabilities.

*Limitations of the study*

This research analyzed problem solving capabilities of LLM-based chatbots for a cross-section of spatial tasks from a variety of categories. Some of these categories are inherently spatial (e.g. spatial literacy, GIS concepts, spatial reasoning) and related tasks were already examined for chatbots earlier. As opposed to this, other categories (e.g., function interpretation, coding) are more general and can be applied to many (including aspatial) disciplines. For these categories we identified or constructed spatial tasks for evaluation. The list of categories examined in this study is not exhaustive since previous studies have analyzed tasks from other spatial categories before, including toponym recognition (Mai et al., acc.), geospatial knowledge extraction (e.g. population density) for a given location (Manvi et al., 2023), and map reading (J. Xu & Tao, 2024). To keep the presented study concise, additional spatial task categories may be considered in follow-up chatbot comparison studies in the future. In addition, this study did not consider complex questions previously described as *indirect Q*A (Scheider et al., 2021). These questions require determining appropriate data sources and tools as well as computation algorithms for an answer. Currently, there is no universally accepted approach to handle these questions in GeoQA. While there are efforts to overcome these limitations, e.g. by applying a grammar to interpret geo-analytical questions and therefore enable machines to solve these kinds of tasks (H. Xu, Nyamsuren, Scheider, & Top, 2023), more research is needed to implement these in the context of LLMs and chatbots.

The current trend of large models is to be multimodal. Spatial tasks and GIS related questions may also include diagrams, flowcharts, and maps (Bolstad & Manson, 2022). Therefore, for future work, some tasks presented in this study may be modified to be presented as multimodal expressions, such as stating a textual question together with a chart, map, or image (J. Xu & Tao, 2024).

Whereas a zero-shot prompt directly instructs the model to perform a task without any additional examples fine-tuning allows developers to train the model on a small dataset for a specific target application or domain (e.g. legal, medical), leading to more accurate and relevant responses (Kasneci et al., 2023). The construction of the prompt, i.e., specialized pretraining through provision of context or personalization, can have an impact on the obtained results as well (Kocoń et al., 2023). For example, few-shot prompting enables in-context learning where the input provides demonstrations in the prompt to steer the model to better performance (Brown et al., 2020). The Chain-of-Thought (CoT) prompt (Wei et al., 2022) includes intermediate steps of reasoning within the prompt besides the task input and output and thus requires the LLM to reason about its strategy before taking actions. It has been found to shown to be effective for tasks that require multiple steps of reasoning and decision-making (F. Li et al., 2024). The Tree of Thoughts (ToT) framework (Yao, Yu, et al., 2023) generalizes over the CoT approach to perform deliberate decision making by considering multiple different reasoning paths and self-evaluating choices to decide the next step. The Graph of Thoughts (GoT) framework provides further prompting advancements through modelling the information generated by an LLM as an arbitrary graph, where LLM thoughts are vertices, and edges correspond to dependencies between these vertices (Besta et al., 2024). The Re-Act prompt (Aghzal et al., 2024; Yao, Zhao, et al., 2023) synergizes reasoning traces (e.g. to track and update action plans, correct a mistaken trajectory) and task-specific actions (e.g. to interface with external sources, such as knowledge bases or environments) in LLMs. A self-checking prompt instructing the LLM to check its answer allows the LLM to identify and rectify its previous mistakes (Kevian et al., 2024; Kojima, Gu, Reid, Matsuo, & Iwasawa, 2022). While in this study a zero-shot prompt approach was applied, future work will consider prompt engineering for tasks that were not successfully completed or answered in this study.

**Conclusions**

LLMs, especially OpenAI's ChatGPT platforms, already demonstrated a significant addition to recent advancements of geo-data analysis methods, sources, and tools, such as crowd-sourcing, citizen science, social media, GIS cloud computing, or blockchain technology (Hochmair, Navratil, & Huang, 2023). The aim of the presented comparison study was to enhance our understanding of the current chatbot ecosystem and its performance in different spatial task categories. The ongoing enhancement of foundation models will improve performance in the multimodal nature of GIS, mapping, and other spatial tasks due to their improved capabilities to reason over various types of geospatial data (e.g., image, video, text, sound) through geospatial alignments (Iyer, Ganguli, & Pandey, 2023; Mai et al., acc.) and the use of geospatial knowledge graphs (Gao et al., 2023; Mai et al., 2022). Evaluation and comparison of new AI technologies and data structures for solving geospatial tasks is part of future work.

**Data availability statement**

The complete set of tasks assigned to chatbots in this study and their responses can be downloaded from https://doi.org/10.6084/m9.figshare.25903729

**Disclosure statement**

No potential conflict of interest was reported by the authors.

**Tables**

Table 1. Correctness of responses in seven task categories, separated by chatbot.

| Category | GPT-4 | Gemini | Claude-3 | Copilot | Category total |
|---|---|---|---|---|---|
| Spatial literacy (SL) | **15/16 (93.8%)** | 11/16 (68.8%) | 13/16 (81.3%) | 13/16 (81.3%) | 52/64 (81.3%) |
| GIS concepts (GIS) | **10/10 (100.0%)** | 9/10 (90.0%) | 9/10 (90.0%) | **10/10 (100.0%)** | 38/40 (95.0%) |
| Mapping (Map) | **6/10 (60.0%)** | 0/10 (0.0%) | 2/10 (20.0%) | 2/10 (20.0%) | 10/40 (25.0%) |
| Function interpretation (Func) | **10/10 (100.0%)** | 7/10 (70.0%) | 8/10 (80.0%) | 8/10 (80.0%) | 33/40 (82.5%) |
| Code explanation (ExpC) | **10/10 (100.0%)** | 8/10 (80.0%) | **10/10 (100.0%)** | **10/10 (100.0%)** | 38/40 (95.0%) |
| Coding (Code) | 5/10 (50.0%) | 3/10 (30.0%) | 4/10 (40.0%) | **7/10 (70.0%)** | 19/40 (47.5%) |
| Spatial reasoning (SR) | 2/10 (20.0%) | **4/10 (40.0%)** | 3/10 (30.0%) | **4/10 (40.0%)** | 13/40 (32.5%) |
| Chatbot total | **58/76 (76.3%)** | 42/76 (55.3%) | 49/76 (64.5%) | 54/76 (71.1%) | |

Table 2. Results of pairwise chi-square post-hoc tests examining the significance of differences in percent correct between task categories

| Category 1 | Category 2 | df | p | $X^2$ |
|---|---|---|---|---|
| Spatial literacy | Mapping | 1 | 4.20E-08 | 30.1 |
| Spatial literacy | Coding | 1 | 7.20E-04 | 11.4 |
| Spatial literacy | Spatial reasoning | 1 | 1.69E-06 | 22.9 |
| GIS concepts | Mapping | 1 | 7.19E-10 | 38.0 |
| GIS concepts | Coding | 1 | 8.73E-06 | 19.8 |
| GIS concepts | Spatial reasoning | 1 | 2.38E-08 | 31.2 |
| Function interpretation | Mapping | 1 | 8.00E-07 | 24.3 |
| Function interpretation | Coding | 1 | 0.0023 | 9.3 |
| Function interpretation | Spatial reasoning | 1 | 1.73E-05 | 18.5 |
| Code explanation | Mapping | 1 | 7.19E-10 | 38.0 |
| Code explanation | Coding | 1 | 8.73E-06 | 19.8 |
| Code explanation | Spatial reasoning | 1 | 2.38E-08 | 31.2 |

Table 3. Word count of responses to spatial literacy questions (n=16) and GIS concepts questions (n=10) in four chatbots; standard deviation shown in parentheses. Small letters indicate pairs of chatbots with significant differences in word count identified in post-hoc tests.

|  | GPT-4 | Gemini | Claude-3 | Copilot | All |
|---|---|---|---|---|---|
| **Spatial literacy** | | | | | |
| Mean (SD) | 66.3 (54.9)[a] | 68.1 (63.2) | 103.0 (47.2)[a] | 74.4 (29.0) | 78.1 (51.3) |
| Median | 45.8 | 50.2 | 116.0 | 73.5 | 59.3 |
| **GIS concepts** | | | | | |
| Mean (SD) | 84.1 (25.4)[b] | 84.9 (37.8)[c] | 172.4 (70.8)[bc] | 105.7 (35.3) | 111.8 (57.1) |
| Median | 82.5 | 76.5 | 177.3 | 95.8 | 90.0 |

Table 4. Chatbots and their underlying model architectures

| Chatbot | Underlying LLM | Architecture |
|---|---|---|
| ChatGPT-4 | GPT-4 | GPT |
| Gemini | Gemini[*] | MoE |
| Claude-3 | Claude-3 | GPT |
| Copilot | Microsoft Prometheus[**] | GPT |

[*] Gemini is multimodal by design.
[**] Microsoft's Prometheus model is based on OpenAI's GPT-4 complemented by Microsoft ecosystem, including search results (Microsoft, 2023)

**Figure captions**

Figure 8. Percentage of correct task completion grouped by chatbot (a) and task category (b). Pairs of tasks with significant differences in percent correct are connected through horizontal bars in (b).

Figure 9. Consistency of responses for task categories, grouped by chatbot (a) and task category (b).

Figure 10. Selected results of mapping tasks. GPT-4: Mapbox map with a line from Vienna to Munich (a); GPT-4: map showing the locations of the 5 largest cities in the U.S. using matplotlib library (b); GPT-4: scatter plot showing the locations of the 5 largest cities in the U.S. using pandas library (c); Claude-3: World map with orthographic map projection using matplotlib library (d); Claude-3: World map using R tmap package (e); Copilot: World map using R tidyverse package (f)

Figure 11. Selected results of coding tasks: Complete R code to plot three polygons in GPT-4 (a) and Claude-3 and Copilot (b); Complete R code to plot a raster map using indirect distance weighting in Copilot (c).